\documentclass[11pt]{article}
\usepackage{moriond,epsfig}
\usepackage{amssymb,amsmath}

\def\be{\begin{equation}}
\def\ee{\end{equation}}
\def\bea{\begin{eqnarray}}
\def\eea{\end{eqnarray}}

\begin{document}
\hfill {\tt CERN-PH-TH/2012-122}

\vspace*{4cm}
\title{Direct and indirect searches for New Physics}

\author{ F. Mahmoudi }

\address{CERN Theory Division, Physics Department, CH-1211 Geneva 23, Switzerland}

\address{Clermont Universit{\'e}, Universit\'e Blaise Pascal, CNRS/IN2P3,\\
LPC, BP 10448, 63000 Clermont-Ferrand, France}

\maketitle\abstracts{
An overview of the indirect constraints from flavour physics on supersymmetric models is presented.
We study in particular constraints from $B_s\to\mu^+\mu^-$ and $B\to K^*\mu^+\mu^-$, emphasising on the new LHCb results. We show that these rare transitions provide valuable information in the search for new physics and are complementary to the direct searches.
}

%%%%%%%%%%%%%%%%%%%%%%%%%%%%%%%%%%%
\section{Introduction}
In addition to direct searches for new physics signals, indirect searches play an important and complementary role in the quest for physics beyond the Standard Model (SM). The most commonly used indirect constraints originate from flavour physics observables, cosmological data and dark matter relic density, electroweak precision tests and anomalous magnetic moment of the muon. 
Precise experimental measurements and theoretical predictions have been achieved for the $B$ meson systems in the past decade and stringent constraints due to sizeable new physics contributions to many observables can be obtained \cite{newphysics}.
In the following, we present an overview of the most constraining flavour physics observables for supersymmetry (SUSY), with an emphasis on the recent LHCb results.
The latest limit on BR($B_s\to\mu^+\mu^-$), being very close to the SM prediction, constrains strongly the large $\tan\beta$ regime and the various angular observables from $B\to K^*\mu^+\mu^-$ decay can provide complementary information in particular for intermediate $\tan\beta$ values. 
We highlight here some of the implications for several SUSY scenarios and show that
these indirect constraints can be superior to those which are derived from direct searches for SUSY particles in some regions of the parameter space.
%
%%%%%%%%%%%%%%%%%%%%%%%%%%%%%%%%%%%
\section{Flavour observables}
\subsection{Framework}
The effective Hamiltonian describing the $b \to s$ transitions has the following generic structure:
\begin{equation}
{\cal H}_{\rm eff}  =  -\frac{4G_{F}}{\sqrt{2}} V_{tb} V_{ts}^{*} \, \Bigl(\,\sum_{i=1\cdots10} \bigl(C_{i}(\mu) O_i(\mu)+C'_{i}(\mu) O'_i(\mu)\bigr)\Bigr)\;,
\end{equation}
where $O_i(\mu)$ are the relevant operators and $C_i(\mu)$ the corresponding Wilson coefficients evaluated at the scale $\mu$ which encode short-distance physics.
The primed operators are chirality flipped compared to the non-primed operators, and they are highly suppressed in the SM.
Contributions from physics beyond the SM to the observables can be described by the modification of Wilson coefficients or by the addition of new operators.
The most relevant operators in rare radiative, semileptonic and leptonic $B$ decays are:
\begin{align}
\label{physical_basis}
O_1& =  (\bar{s} \gamma_{\mu} T^a P_L c) (\bar{c} \gamma^{\mu} T^a P_L b)\;,\quad
&O_2& = (\bar{s} \gamma_{\mu} P_L c) (\bar{c} \gamma^{\mu} P_L b)\;,\nonumber\\
O_3& =  (\bar{s} \gamma_{\mu} P_L b) \sum_q (\bar{q} \gamma^{\mu} q)\;,\quad
&O_4& = (\bar{s} \gamma_{\mu} T^a P_L b) \sum_q (\bar{q} \gamma^{\mu} T^a q)\;,\nonumber\\
O_5& =  (\bar{s} \gamma_{\mu_1} \gamma_{\mu_2} \gamma_{\mu_3} P_L b) 
                  \sum_q (\bar{q} \gamma^{\mu_1} \gamma^{\mu_2} \gamma^{\mu_3} q)\;,\quad
&O_6& = (\bar{s} \gamma_{\mu_1} \gamma_{\mu_2} \gamma_{\mu_3} T^a P_L b) 
                  \sum_q (\bar{q} \gamma^{\mu_1} \gamma^{\mu_2} \gamma^{\mu_3} T^a q)\;,\nonumber\\
O_7& = \frac{e}{(4\pi)^2} m_b (\overline{s} \sigma^{\mu\nu} P_R b) F_{\mu\nu} \;, \quad
&O_8& = \frac{g}{(4\pi)^2} m_b (\bar{s} \sigma^{\mu \nu} T^a P_R b) G_{\mu \nu}^a \;,
\\ \nonumber
O_9& =  \frac{e^2}{(4\pi)^2} (\overline{s} \gamma^\mu P_L b) (\bar{\ell} \gamma_\mu \ell) \;, \quad
&O_{10}& =  \frac{e^2}{(4\pi)^2} (\overline{s} \gamma^\mu P_L b) (\bar{\ell} \gamma_\mu \gamma_5 \ell)\;, \\ \nonumber
Q_1& = \frac{e^2}{(4\pi)^2} (\bar{s} P_R b)(\bar{\ell}\,\ell)\;, \quad
&Q_2& =  \frac{e^2}{(4\pi)^2} (\bar{s} P_R b)(\bar{\ell}\gamma_5 \ell)\;,
\end{align}
where $Q_1$ and $Q_2$ are the scalar and pseudo-scalar operators arising in new physics scenarios.

The Wilson coefficients $C_i(\mu)$ are calculated at scale $\mu \sim \mathcal{O}(M_W)$ by requiring matching between the effective and full theories. They can be expanded perturbatively:
\begin{equation}
C_i(\mu) = C^{(0)}_i(\mu) + \frac{\alpha_s(\mu)}{4 \pi} C^{(1)}_i(\mu) + \cdots
\end{equation}
and are subsequently evolved to scale $\mu \sim \mathcal{O}(m_b)$ at which they can be used to calculate the flavour observables, using the renormalisation group equations:
\begin{equation}
\mu \frac{d}{d \mu} C_i(\mu) = C_j(\mu) \gamma_{ji}(\mu)
\end{equation}
driven by the anomalous dimension matrix $\hat{\gamma}(\mu)$: 
\begin{equation}
\hat{\gamma}(\mu) = \frac{\alpha_s (\mu)}{4 \pi} \hat{\gamma}^{(0)}  
                        + \frac{\alpha_s^2(\mu)}{(4 \pi)^2} \hat{\gamma}^{(1)} + \cdots 
\end{equation}
which are known to high accuracy. A review on effective methods is given in \cite{Buras:1998raa} and the analytical expressions for the Wilson coefficients and the renormalisation group equations can be found in~\cite{Mahmoudi:2008tp}.
%
%%%%%%%%%%%%%%%%%%%%%%%%%%%%%%%%%%%
\subsection{Observables}
The rare decays $B_s\to\mu^+\mu^-$ and $B\to K^*\mu^+\mu^-$ deserve special attention as new results have been recently announced by the LHCb collaboration using an integrated luminosity of 1~fb$^{-1}$. In particular, a stringent 95\% C.L. limit on the branching ratio BR$(B_s\to\mu^+\mu^-) < 4.5\times 10^{-9}$ has been obtained \cite{Aaij:2012ac}. 
In terms of Wilson coefficients, this branching ratio is expressed as \cite{Mahmoudi:2008tp,Bobeth:2001sq}:
\begin{eqnarray}
  \label{eq:Bs2mm_formula}
\mbox{BR}(B_s\to\mu^+\mu^-)&=&\frac{G_F^2 \alpha^2}{64\pi^2}f_{B_s}^2
m_{B_s}^3 |V_{tb}V_{ts}^*|^2\tau_{B_s}\sqrt{1-\frac{4m_\mu^2}{m_{B_s}^2}}\\
&&\times\left\{\left(1-\frac{4m_\mu^2}{m_{B_s}^2}\right)
  |C_{Q_1}-C'_{Q_1}|^2+\left|(C_{Q_2}-C'_{Q_2})+2(C_{10}-C'_{10})\frac{m_\mu}{m_{B_s}}\right|^2\right\}\,.\nonumber  
\end{eqnarray}
In the Standard Model, only $C_{10}$ is non-vanishing and gets its largest contributions from $Z$ penguin and box diagrams. With the input parameters of \cite{Mahmoudi:2012xx} we obtain BR$(B_s\to\mu^+\mu^-)_{SM}=(3.53\pm 0.38)\times 10^{-9}$. The latest experimental limit thus severely restricts the room for new physics.

The decay $B\to K^*\mu^+\mu^-$ on the other hand provides a variety of complementary observables
as it gives access to angular distributions in addition to the differential branching fraction. 
The differential decay distribution of the $\bar B ^0 \to \bar K ^*(\to K^- \pi^+ ) \mu^+ \mu^-$ decay 
can be written as a function of three angles $\theta_l$, $\theta_{K^*}$, $\phi$ and 
the invariant dilepton mass squared ($q^2$) \cite{Kruger:2005ep,Altmannshofer:2008dz}:
\begin{equation}\label{eq:diffAD}
  d^4\Gamma = \frac{9}{32\pi} J(q^2, \theta_l, \theta_{K^*}, \phi)\, dq^2\, d\cos\theta_l\, d\cos\theta_{K^*}\, d\phi \;.
\end{equation}
The angular dependence of $J(q^2, \theta_l, \theta_{K^*}, \phi)$ are then expanded in terms of the angular coefficients $J_i$ which are functions of $q^2$ and can be described in terms of the transversity amplitudes and form factors \cite{Beneke:2001at,Beneke:2004dp}.
Integrating Eq. \ref{eq:diffAD} over all angles, the dilepton mass distribution is obtained in terms of the angular coefficients \cite{Altmannshofer:2008dz,Bobeth:2008ij}:
\begin{equation}
\frac{d\Gamma}{dq^2} = \frac{3}{4} \bigg( J_1 - \frac{J_2}{3} \bigg)\;.
\label{eq:dBR}
\end{equation}
The forward-backward asymmetry $A_{FB}$, which benefits from reduced theoretical uncertainty, is defined as:
\begin{equation}
A_{\rm FB}(q^2)  \equiv
     \left[\int_{-1}^0 - \int_{0}^1 \right] d\cos\theta_l\, 
          \frac{d^2\Gamma}{dq^2 \, d\cos\theta_l} \Bigg/\frac{d\Gamma}{dq^2}
          =  -\frac{3}{8} J_6 \Bigg/ \frac{d\Gamma}{dq^2}\; .
\label{eq:AFB}
\end{equation}
Another clean observable is the zero--crossing of the forward-backward asymmetry ($q_0^2$) for which the form factors cancel out at leading order. $q_0^2$ depends on the relative sign of $C_7$ and $C_9$ and its measurement allow to remove the sign ambiguity. 

The longitudinal polarisation fraction $F_L$ can also be constructed as the ratio of the transversity amplitudes and contains less theoretical uncertainty from the form factors. It reads:
\begin{equation}
 F_L(s) = \frac{-J_2^c}{d\Gamma / dq^2}\;.
\end{equation}
The SM predictions and experimental values for these observables are given in Table~\ref{tab:experiment}.
\begin{table}[!t]
\begin{center}
\begin{tabular}{|l|l|l|l|l|l|l|}\hline 
  Observable                                                                & SM prediction & Experiment       \\ \hline \hline
  $10^7 \mbox{GeV}^2 \times \langle dBR/dq^2\; (B \to K^* \mu^+ \mu^-) \rangle_{[1,6]}$ & $0.47 \pm 0.27 $        & $0.42 \pm 0.04 \pm 0.04$   \\ \hline
  $\langle A_{FB}(B \to K^* \mu^+ \mu^-) \rangle_{[1,6]}$         & $-0.06 \pm 0.05 $        & $-0.18 ^{+0.06+0.01}_{-0.06-0.01}$   \\ \hline
  $\langle F_{L}(B \to K^* \mu^+ \mu^-) \rangle_{[1,6]}$          & $0.71 \pm 0.13 $        & $0.66 ^{+0.06+0.04}_{-0.06-0.03}$   \\ \hline
  $q_0^2 (B \to K^* \mu^+ \mu^-)/\mbox{GeV}^2$      & $4.26 ^{+0.36}_{-0.34} $        & $4.9  ^{+1.1}_{-1.3}$   \\ \hline
 \end{tabular}
\caption{SM predictions and experimental values of $B\to K^*\mu^+\mu^-$ observables $^6$. %\cite{Mahmoudi:2012xx}.
\label{tab:experiment}}
\end{center}
\end{table}

Another observable which is rather independent of hadronic input parameters is the isospin asymmetry which arises from the annihilation diagrams and depends on the charge of the spectator quark. The isospin asymmetry is defined as \cite{Feldmann:2002iw}
\begin{eqnarray}
\frac{dA_I}{dq^2} &= & 
\frac{ d\Gamma[B^0\to K^{\ast0}\mu^+ \mu^-]/dq^2 -
d\Gamma[B^\pm\to  K^{\ast\pm}\mu^+ \mu^-]/dq^2}
{ d\Gamma[ B^0\to K^{\ast0}\mu^+ \mu^-]/dq^2 +
d\Gamma[B^\pm \to  K^{\ast\pm}\mu^+ \mu^-]/dq^2} \ .
\label{isospin}
\end{eqnarray}
In the SM, $dA_I/dq^2$ is at the percent level. 

The decay $B\to K^*\mu^+\mu^-$ gives access to many other observables such as transverse amplitudes, which are not yet measured but could be of interest in the near future.

In addition to the above observables, $B\to X_s \gamma$, $B\to \tau \nu$, $B \to D \tau \nu_\tau$, $B \to X_s \mu^+ \mu^-$ and $D_s \to \tau \nu_\tau$ are also very sensitive to SUSY as discussed in \cite{newphysics}.
%
%%%%%%%%%%%%%%%%%%%%%%%%%%%%%%%%%%%
\section{Implications for supersymmetry}
To illustrate the impact of the flavour observables and in particular $B_s\to \mu^+\mu^-$ and $B\to K^*\mu^+\mu^-$, we consider several MSSM scenarios, and compare the resulting constraints to the direct search limits.
\begin{figure}[!t]
\begin{center}
\includegraphics[width=7.5cm]{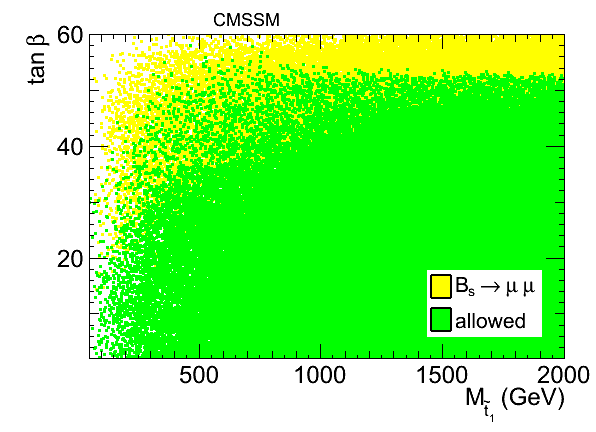}\includegraphics[width=7.5cm]{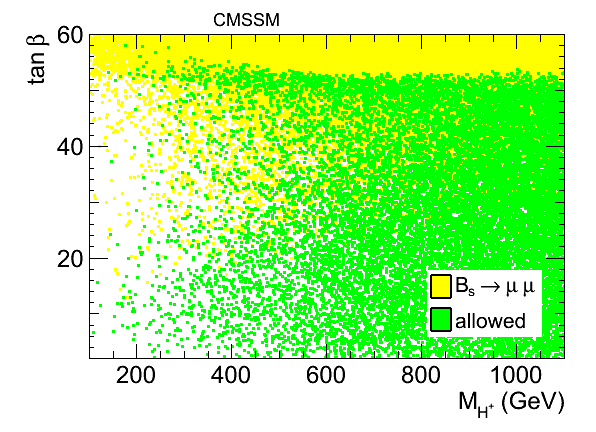}
\includegraphics[width=7.5cm]{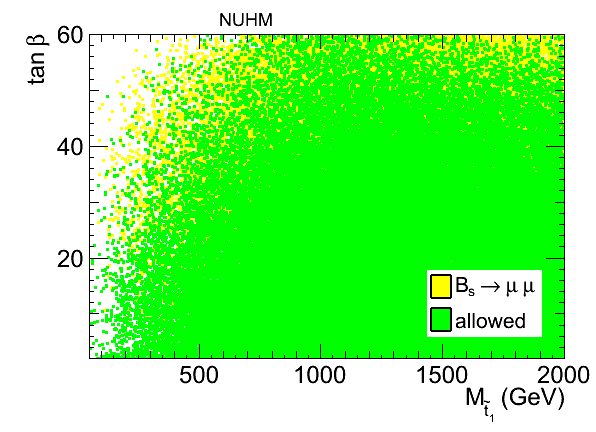}\includegraphics[width=7.5cm]{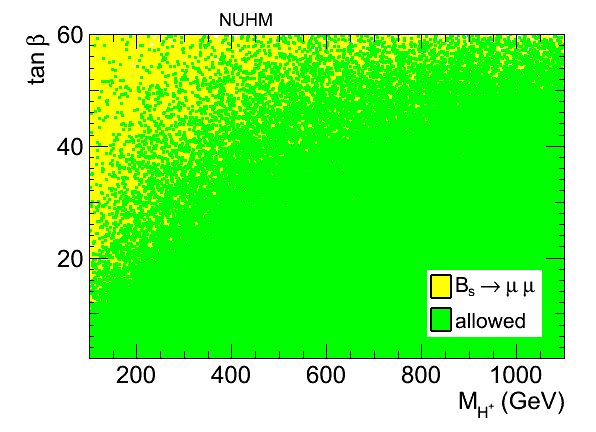}
\includegraphics[width=7.5cm]{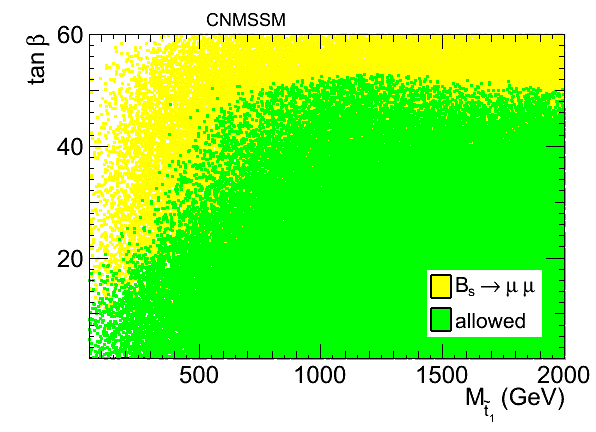}\includegraphics[width=7.5cm]{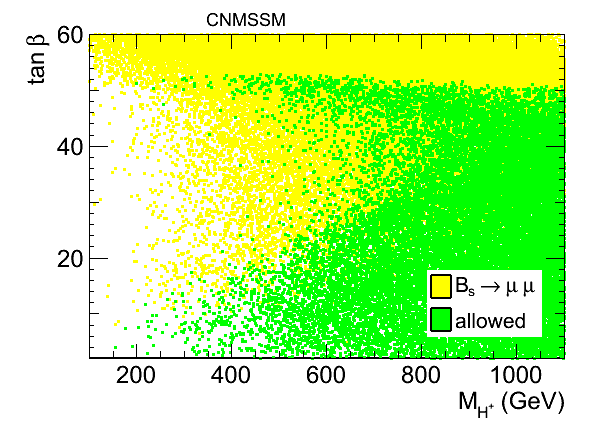}
\caption{Constraints from BR($B_s\to \mu^+\mu^-$) in the CMSSM (upper panel), NUHM (central panel) and CNMSSM (lower panel) in the plane ($M_{\tilde t_1}$, $\tan\beta$) in the left and ($M_{H^{\pm}}$, $\tan\beta$) in the right, with the allowed points displayed in the foreground.\label{fig:Bsmumu}}
\end{center}
\end{figure}
\begin{figure}[!t]
\begin{center}
\includegraphics[width=7.5cm]{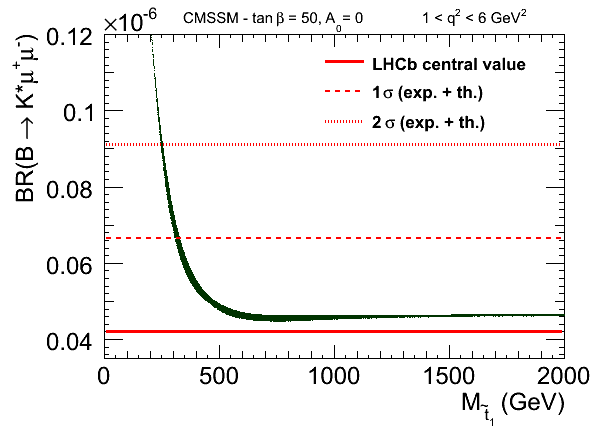}\quad\includegraphics[width=7.5cm]{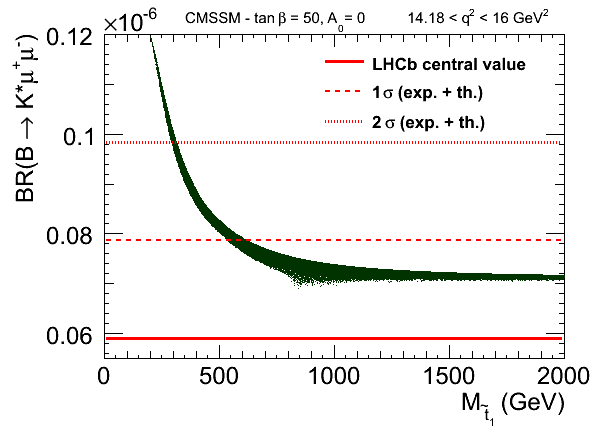}
\includegraphics[width=7.5cm]{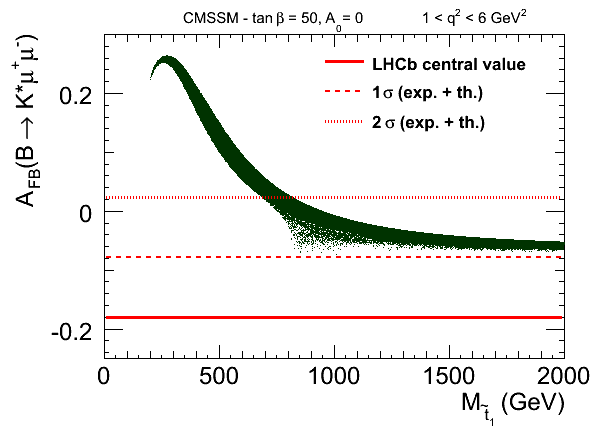}\quad\includegraphics[width=7.5cm]{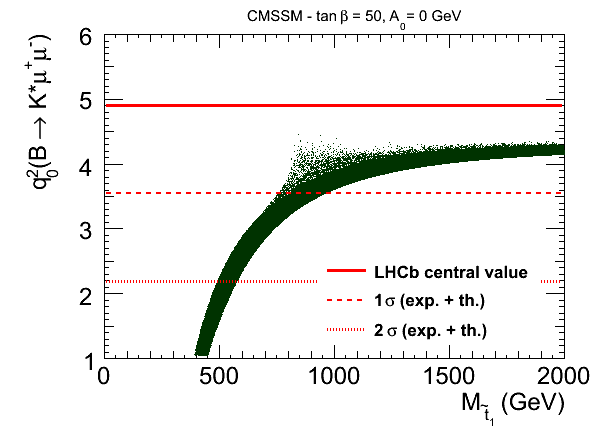}
\includegraphics[width=7.5cm]{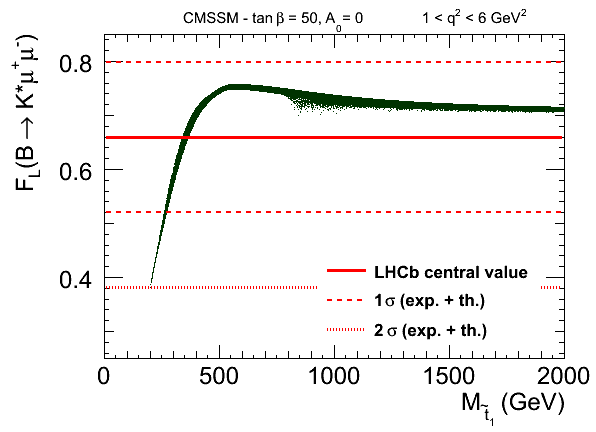}\quad\includegraphics[width=7.5cm]{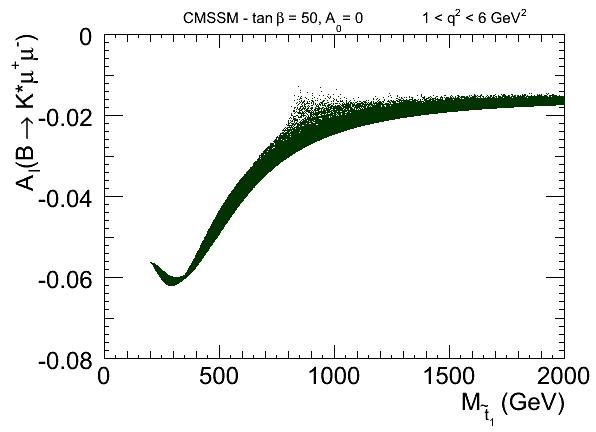}
\caption{SUSY spread of the averaged BR($B\to K^*\mu^+\mu^-$) at low $q^2$ (top left), at high $q^2$ (top right), $A_{FB}(B\to K^*\mu^+\mu^-)$ at low $q^2$ (middle left), zero-crossing of $A_{FB}(B\to K^*\mu^+\mu^-)$ (middle right), $F_L(B\to K^*\mu^+\mu^-)$ at low $q^2$ (bottom left) and $A_I(B\to K^*\mu^+\mu^-)$ at low $q^2$ (bottom right), as a function of the lightest stop mass, in the CMSSM with $\tan\beta$=50 and $A_0=0$. \label{fig:bsll}}
\end{center}
\end{figure}
\begin{figure}[!t]
\begin{center}
\includegraphics[width=7.5cm]{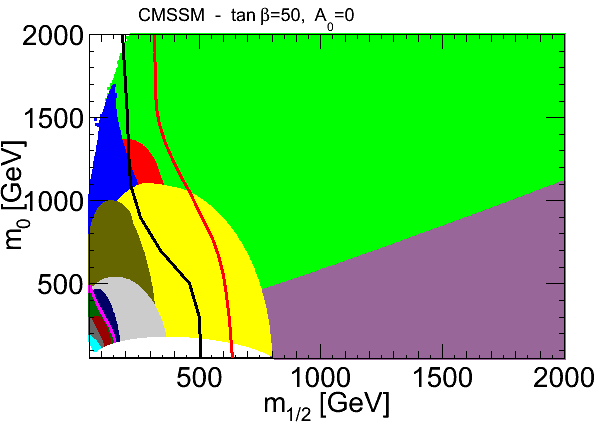}
\includegraphics[width=7.5cm]{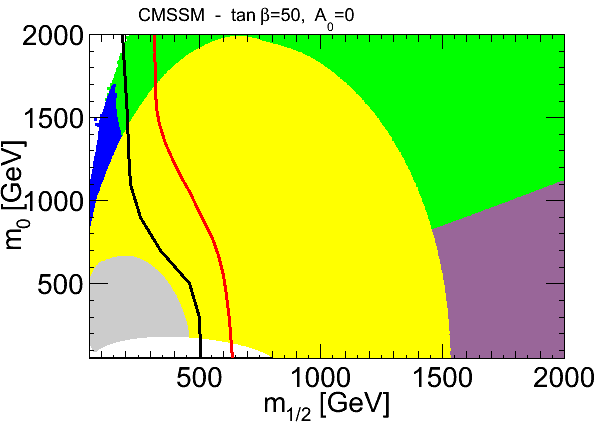}
\caption{Constraints from flavour observables in CMSSM in the plane ($m_{1/2}, m_0$) for $\tan\beta=$ 50 and $A_0=0$, in the left with the 2010 results for BR($B_s\to \mu^+\mu^-$), and in the right with the 2011 results. The black line corresponds to the CMS exclusion limit with 1.1 fb$^{-1}$ of data $^{18}$
%\cite{Chatrchyan:2011zy}
and the red line to the CMS exclusion limit with 4.4 fb$^{-1}$ of data $^{19}$
%\cite{CMS-PAS-SUS-12-005}
. The colour legend is given below.}
\label{fig:comb}
\end{center}
\end{figure}
\begin{figure}[!t]
\begin{center}
\includegraphics[width=7.5cm]{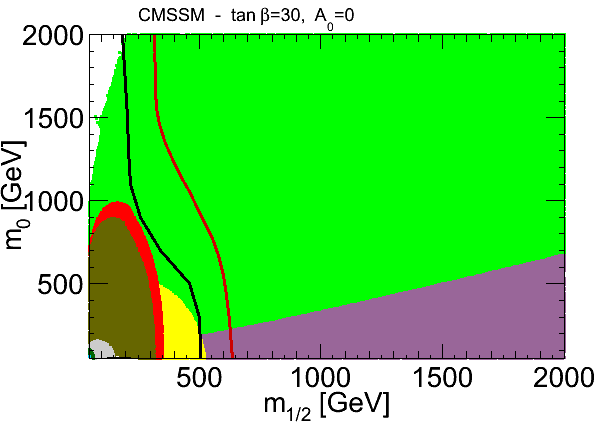}\quad\quad\includegraphics[width=2.5cm]{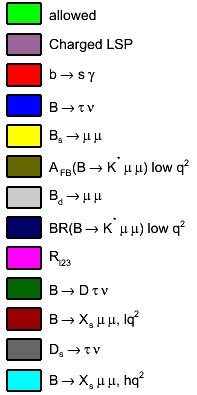}
\caption{Constraints from flavour observables in CMSSM in the plane ($m_{1/2}, m_0$) for $\tan\beta=$ 30 and $A_0=0$.}
  \label{fig:comb30}
\end{center}
\end{figure}

First we study the constraints from BR($B_s\to \mu^+\mu^-$) in the CMSSM, NUHM and CNMSSM by scanning over the relevant parameters as described in \cite{Mahmoudi:2012xx,Akeroyd:2011kd}. For each generated point we calculate the spectrum of SUSY particle masses and couplings using SOFTSUSY \cite{Allanach:2001kg} or NMSSMTOOLS \cite{Ellwanger:2005dv} and compute flavour observables using SuperIso v3.3 \cite{Mahmoudi:2008tp,superiso}.

The constraints are shown in Fig.~\ref{fig:Bsmumu} in the planes ($M_{\tilde t_1}$, $\tan\beta$) and ($M_{H^{\pm}}$, $\tan\beta$). The region most probed by $B_s\to \mu^+\mu^-$ is at large $\tan\beta$ and small $M_{\tilde t_1}$ / $M_{H^{\pm}}$ as can be seen from the figures. Since there are two additional degrees of freedom in NUHM as compared to CMSSM, it is easier for a model point to escape the limits and the constraints are therefore weaker in NUHM. In the CNMSSM, the $B_s\to \mu^+\mu^-$ constraint is similar to the CMSSM case, but slightly stronger.

Next we consider the constraints from $B\to K^*\mu^+\mu^-$ observables.
In order to study the maximal effects we consider $\tan\beta$=50 and investigate the SUSY spread as a function of the lightest stop mass. The results are displayed in Fig.~\ref{fig:bsll} for the averaged differential branching ratio at low and high $q^2$, the forward-backward asymmetry $A_{FB}$, the zero-crossing $q^2_0$ of $A_{FB}$, the longitudinal polarisation $F_L$ and the isospin asymmetry $A_I$. The solid red lines correspond to the LHCb central value, while the dashed and dotted lines represent the 1 and 2$\sigma$ bounds respectively, including both theoretical and experimental errors (added in quadrature).
As can be seen from the figure, $A_{FB}$ is the most constraining observable and excludes $M_{\tilde t_1} \lesssim$ 800 GeV. On the other hand, with the current experimental accuracy \cite{collaboration:2012cq}, the isospin asymmetry does not provide any information on the SUSY parameters.

A comparison between different flavour observables in the plane ($m_{1/2},m_0$) is given in Fig.~\ref{fig:comb}, where we can also see the limits from $B\to X_s \gamma$, $B\to \tau \nu$, $R_{l23}(K\to\mu \nu_\mu)$, $B \to D \tau \nu_\tau$, $B \to X_s \mu^+ \mu^-$ and $D_s \to \tau \nu_\tau$. 
In the left hand side, the combined CMS+LHCb limit from the 2010 data ($1.1 \times 10^{-8}$ at 95\% C.L.) is applied for BR($B_s\to \mu^+\mu^-$), while this limit is updated to the 2011 LHCb result ($4.5 \times 10^{-9}$ at 95\% C.L.) in the right hand side. As can be seen, the recent LHCb limit strongly constrains the CMSSM with large $\tan\beta$.
We also notice that, at large $\tan\beta$, the flavour constraints and in particular $B_s \rightarrow \mu^+ \mu^-$, are superior to those from direct searches. By lowering the value of $\tan\beta$, $B_s\to \mu^+\mu^-$ significantly loses importance compared to direct searches as can be seen in Fig.~\ref{fig:comb30}. On the other hand, $B\to X_s \gamma$ and $B \to K^* \mu^+\mu^-$ related observables and in particular the forward-backward asymmetry lose sensitivity in a less drastic manner and they could play a complementary role in the intermediate $\tan\beta$ regime.

\begin{figure}[!t]
\begin{center}
\includegraphics[width=7.5cm]{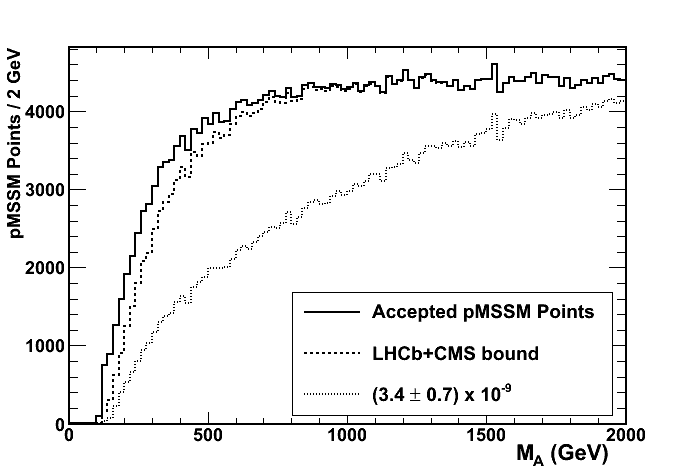}
\includegraphics[width=7.5cm]{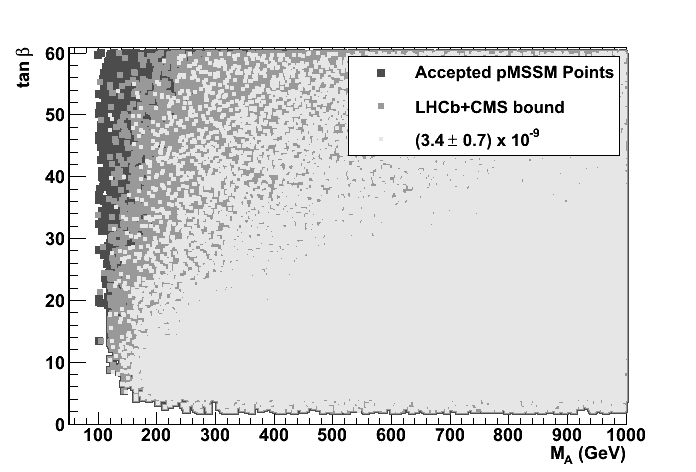}
\caption{Distribution of pMSSM points after the $B_s \rightarrow \mu^+ \mu^-$ constraint
projected on the $M_A$ (left) and ($M_A , \tan \beta$) plane (right) for all accepted pMSSM points 
(medium grey), points not excluded by the combination of the 2010 LHCb and CMS analyses (dark grey) 
and the projection for the points compatible with the measurement of the SM expected branching 
fractions with a 20\% total uncertainty (light grey) $^{21}$.
%\cite{Arbey:2011aa}
}
  \label{fig:pmssm}
\end{center}
\end{figure}

The study in constrained MSSM scenarios is very illustrative and allows to pin down the most important effects in a rather simple framework. However these scenarios are not representative of the full MSSM and by focussing only on the constrained scenarios one may miss some important features. Also the constrained scenarios are already very much squeezed, while this is not the case in more general scenarios. 
To go beyond the constrained scenarios, we consider the phenomenological MSSM (pMSSM) \cite{pmssm}.
This model is the most general CP and R--parity conserving MSSM, assuming MFV at the weak scale and the absence of FCNCs at the tree level. It contains 19 free parameters: 10 sfermion masses, 3 gaugino masses, 3 trilinear couplings and 3 Higgs masses.
To study the pMSSM, we perform flat scans over the parameters as described in \cite{Arbey:2011aa,Arbey:2011un}.
The left panel of Fig.~\ref{fig:pmssm} shows the density of points in function of $M_{A}$ before and after applying the combined 2010 LHCb and CMS limit for  $B_s \to \mu^+ \mu^-$, as well as the projection for an SM--like measurement with an overall 20\% theoretical and experimental uncertainty. As can be seen the density of the allowed pMSSM points is reduced by a factor of 3, in the case of an SM--like measurement. The right panel shows the same distribution in the ($M_{A}$, $\tan\beta$) plane. The region with large $\tan\beta$ and small $M_{A}$ is the most affected one.
%
%%%%%%%%%%%%%%%%%%%%%%%%%%%%%%%%%%%
\vspace*{-0.2cm}
\section{SuperIso program}\vspace*{-0.1cm}
SuperIso \cite{Mahmoudi:2008tp,superiso} is a public C program dedicated mainly to the calculation of flavour physics observables. 
The calculations are done in various models, such as SM, 2HDM, MSSM and NMSSM with minimal flavour violation.
A broad set of flavour physics observables is implemented in SuperIso. This includes the branching ratio of $B \to X_s \gamma$, isospin asymmetry of $B \to K^* \gamma$, branching ratios of $B_s \to \mu^+ \mu^-$, $B_d \to \mu^+ \mu^-$, $B_u \to \tau \nu_\tau$, $B \to D \tau \nu_\tau$, $K \to \mu \nu_\mu$, $D \to \mu \nu_\mu$, $D_s \to \tau \nu_\tau$ and $D_s \to \mu \nu_\mu$. In addition several observables related to $b\to s\ell^+\ell^-$ transitions, such as branching ratios of $B \to X_s \mu^+ \mu^-$ and $B \to K^* \mu^+ \mu^-$, the forward backward asymmetries, the zero-crossings, polarisation fractions of $K^*$, isospin asymmetries, transverse amplitudes, the CP averaged angular coefficients, etc..., have also been included. 

The calculation of the anomalous magnetic moment of the muon is also implemented in the program. SuperIso uses a SUSY Les Houches Accord (SLHA) file \cite{Allanach:2008qq} as input, which can be either generated automatically by the program via a call to a spectrum generator or provided by the user.
The program is able to perform the calculations automatically for different SUSY breaking scenarios. 
An extension of SuperIso including the relic density calculation, SuperIso Relic, is also available publicly \cite{superiso_relic}. 
Finally, in SuperIso we make use of the Flavour Les Houches Accord (FLHA) \cite{Mahmoudi:2010iz}, the newly developed standard for flavour related quantities.
%
%%%%%%%%%%%%%%%%%%%%%%%%%%%%%%%%%%%
\vspace*{-0.1cm}
\section{Conclusions}\vspace*{-0.1cm}
Indirect constraints and in particular those from flavour physics are essential to restrict the new\clearpage \noindent physics parameters as we have seen here. The information obtained from these low energy observables combined with the collider data opens the door to a very rich phenomenology and would help us to step forward toward a deeper understanding of the underlying physics. 
It is clear that with more precise measurements of flavour observables a large part of the supersymmetric parameter space could be disfavoured. In particular large $\tan \beta$ region is strongly affected by $B_s \to \mu^+ \mu^-$. Also, a measurement of BR($B_s \to \mu^+ \mu^-$) lower than the Standard Model prediction would rule out a large variety of supersymmetric models. In addition, $B \to K^* \mu^+\mu^-$ observables play a complementary role specially for smaller $\tan \beta$ values. With reduced theoretical and experimental errors, the exclusion bounds in Fig. \ref{fig:bsll} for example would squeeze leading to important consequences for SUSY parameters. The $B \to K^* \mu^+\mu^-$ decay provides many other clean observables, not yet measured, which could also bring substantial additional information.
%%%%%%%%%%%%%%%%%%%%%%%%%%%%%%%%%%%
\section*{Acknowledgements}
I would like to thank the organisers of the Moriond QCD 2012 for organising a great conference and for inviting me to give this talk.
%
%%%%%%%%%%%%%%%%%%%%%%%%%%%%%%%%%%%

\end{document}